\documentclass[twocolumn, a4paper, notitlepage]{article}
\usepackage{natbib}
\usepackage{graphicx}
\usepackage[colorlinks=true]{hyperref}

\providecommand{\jgr}{J. Geophys. Res.}
\providecommand{\affil}[1]{\textsuperscript{#1}}
\providecommand{\affiliation}[2]{{#1}: {#2}\\}

\pdfoutput=1
\setcitestyle{square}

\makeatletter
\renewcommand{\maketitle}{
    \begin{center}
      \large
      {\LARGE\@title}
      \par\vspace{1ex}
        \@author
	\par\vspace{1ex}
      \@date
    \end{center}
    \@thanks
}
\makeatother
\begin{document}

\title{MARSIS observations of field-aligned irregularities and ducted radio propagation in the Martian ionosphere}

\author{
D. J. Andrews\affil{1},
H. J. Opgenoorth\affil{1},
T. B. Leyser\affil{1},
S. Buchert\affil{1},
N. J. T. Edberg\affil{1},
D. D. Morgan\affil{2},
D. A. Gurnett\affil{2},
A. J. Kopf\affil{2},
K. Fallows\affil{3},
P. Withers\affil{3,4}.\\
\vspace{11pt}
Corresponding author: David J. Andrews, Swedish Institute of Space Physics (Uppsala), Box 537, Uppsala 75121, Sweden.\\
 david.andrews@irfu.se\\
 \vspace{11pt}
 \affiliation{1}{Swedish Institute for Space Physics, Uppsala, Sweden}
 \affiliation{2}{Department of Physics \& Astronomy, University of Iowa, IA, USA}
 \affiliation{3}{Center for Space Physics, Boston University, MA, USA}
 \affiliation{4}{Department of Astronomy, Boston University, MA, USA}
}

\date{\vspace{11pt}
August 2018. Preprint accepted for publication in \\J. Geophys. Res. (Space Physics).\\}

\onecolumn

\maketitle

\begin{abstract}

Knowledge of Mars's ionosphere has been significantly advanced in recent years by observations from Mars Express (MEX) and lately MAVEN.
A topic of particular interest are the interactions between the planet's ionospheric plasma and its highly structured crustal magnetic fields, and how these lead to the redistribution of plasma and affect the propagation of radio waves in the system.
In this paper, we elucidate a possible relationship between two anomalous radar signatures previously reported in observations from the MARSIS instrument on MEX.
Relatively uncommon observations of localized, extreme increases in the ionospheric peak density in regions of radial (cusp-like) magnetic fields and spread-echo radar signatures are shown to be coincident with ducting of the same radar pulses at higher altitudes on the same field lines.
We suggest that these two observations are both caused by a high electric field (perpendicular to $\mathbf{B}$) having distinctly different effects in two altitude regimes.
At lower altitudes, where ions are demagnetized and electrons magnetized, and recombination dominantes, a high electric field causes irregularities, plasma turbulence, electron heating, slower recombination and ultimately enhanced plasma densities.
However, at higher altitudes, where both ions and electrons are magnetized and atomic oxygen ions cannot recombine directly, the high electric field instead causes frictional heating, a faster production of molecular ions by charge exchange, and so a density decrease.
The latter enables ducting of radar pulses on closed field lines, in an analogous fashion to inter-hemispheric ducting in the Earth's ionosphere.

Key points:
\begin{itemize}
  \item MARSIS on MEX observed both echoes from distant ionospheric irregularities as well as locally ducted echoes in close succession
\item Both effects are consistent with the presence of extended, small-scale field-aligned density irregularities
\item We suggest that these may be the direct result of strong electric fields present at ionospheric altitudes
\end{itemize}
\end{abstract}
\twocolumn

\section{Introduction}

A diverse family of plasma instabilities are known to act in the Earth's ionosphere, producing a range of ionospheric irregularities with many different characteristics~\citep[see e.g.][]{fejer80a}.
While some are of purely academic interest, others, such as equatorial `spread-F' can also impair various radio communications.
Despite the generally weaker magnetic field of Mars, many of the same instabilities can be expected to act and modify the ionosphere on a range of spatial and temporal scales.
In this paper, we provide further analysis of recently published observations of Martian ionospheric irregularities, linking enhanced plasma densities at low (below $\sim$200~km) altitudes with ducted radio propagation at higher altitudes.

The ionosphere of Mars, formed principally through the photo ionization of atmospheric CO$_2$, is structured by variations in the ionization rates, neutral winds, solar wind interaction, and the structure of crustal magnetic fields at ionospheric altitudes (see e.g.\ recent reviews by~\cite{witasse02a},  \cite{nagy04a}, \cite{withers09a}).
Observations made by the Mars Global Surveyor (MGS), Mars Express (MEX), and most recently the Mars Atmosphere and Volatile EvolutioN (MAVEN) missions have shown that the crustal fields are responsible for establishing both large-scale changes in the absolute ionospheric plasma density \citep[e.g.][]{nilsson11a, dubinin12a, andrews13a}, small-scale variability~\citep[e.g.][]{brain07a, gurnett10a, andrews15b}, and stable ionospheric ``upwellings''~\citep{gurnett05a, duru06a, andrews14a}.
Crustal fields also either control or at least modulate the character of Martian auroral emissions~\citep[e.g.][]{bertaux05a, brain06b}.

While the combined lack of a planetary dynamo field and the presence of intense crustal fields is unique to Mars within the solar system, the resulting magnetic morphologies at ionospheric altitudes bear some similarities to both the equatorial Earth ionosphere with near horizontal fields, and the auroral or cusp-like ionosphere with near vertical fields.
In contrast to Earth, the inclination of Martian magnetic field can reverse on very small length scales ($\sim$100~km), comparable to the typical ionospheric ion gyroradius.
The electron cyclotron frequency $f_{ce}$ typically remains several orders of magnitude smaller than the electron plasma frequency $f_{pe}$ throughout the Martian ionosphere, such that radio wave propagation is not significantly affected by the magnetic field itself.

In this paper we present data from the Mars Advanced Radar for Sub-Surface and Ionospheric Sounding (MARSIS) instrument, the main antenna of which is a 40~m tip-to-tip dipole.
When operated in active ionospheric sounding (AIS) mode~\citep{picardi04b, gurnett05a} discrete pulses are transmitted at stepped central frequencies between $\sim$100~kHz and $\sim$5.5~MHz.
The part of the pulse propagating in the nadir direction is specularly reflected from the Martian ionospheric plasma at an altitude where the sounding frequency $f=f_{pe}$.
A reflecting structure, such as an ionospheric layer, forms a trace in the reflected signal when plotted as a function of sounding frequency and delay time - a so-called ionogram.
The elapsed time between transmission of the pulse and its receipt yields a measure of the distance to the reflection point, although a full numerical inversion must be performed to remove the effect of the dispersion of the pulse as it propagates through the plasma~\citep{morgan13a}.
The principal purpose of this mode of operation is therefore to obtain vertical profiles of plasma density from the altitude of the spacecraft down to the peak of the ionospheric plasma density.

Close to the terminator and in regions of strong crustal fields, Mars's ionosphere departs significantly from the ideal case of horizontally stratified medium.
This can give rise to reflected signals in the MARSIS data that arrive at the radar from off-nadir directions, i.e. received at oblique incidence.
Many such oblique echoes have been shown to be associated with reflection from stable, large-scale ionospheric upwellings in regions of radial (vertical) crustal fields~\citep{gurnett05a, duru06a}.
However, as MARSIS does not discriminate between the arrival directions of signals received, the inference of the source of these oblique echoes is made in relation to models of the crustal field at and in the vicinity of the spacecraft.
Other ionospheric structures can also give rise to oblique echoes, for example horizontal variations in plasma density at the terminator~\citep{duru10a}.

Comparable top- and bottom-side (ground based) ionospheric radar sounders have been routinely operated at Earth for many decades.
Spread-F was found to be a common feature of Earth's F-region ionosphere, in which an otherwise sharp reflected radar signal is spread over a much larger apparent range, indicating reflections received from a disturbed plasma in contrast to a stably stratified layer.
A connection between instances of spread-F radar signatures and localized ionospheric plasma density depletions (sometimes referred to as `plumes') often detected nearby was suggested in several studies in the late 1970s~\citep{kelley76a, woodman76a, mcclure77a}.
Spread-F and plumes are typically observed in the hours after local sunset, triggered by a strengthening eastward electric field just around sunset, the pre-reversal enhancement~\citep[e.g.][]{woodman76a}.
These under-dense plumes, formed below the ionospheric peak in regions of near-horizontal magnetic field, grow and propagate vertically through the Raleigh-Taylor instability, as first suggested by~\cite{dungey56a}.
As the plumes rise to higher altitudes they perturb the surrounding ionosphere, forming the irregularities that are responsible for the observed radar spread-F.
The resulting density depletions extend along the horizontal fields, and can reach a stable situation in the topside ionosphere.
Space-borne radar experiments were then found to occasionally pass through these plumes, and observe the ducted propagation of radar signals along the field-aligned density cavities~\citep[e.g.,][]{muldrew69a, dyson78a, platt89a}.
These ducts can extend over large distances, often leading to the reception of reflections from the conjugate hemisphere of the planet, in addition to the typical reflections from nadir incidence.
Propagation of sounding pulses within these ducts leads to rather unusual traces in an ionogram, with more complex appearance than the reflection from a nominal ionosphere.
Radar spread-F is also routinely associated with the presence of ducted echoes~\citep{muldrew69b}.


Ducted propagation of radio waves was first identified in the Martian  ionosphere by~\cite{zhang16a}.
In a survey of $\sim$8 years of data from MARSIS, seven ducted propagation events were found by manual inspection of the data, each of which occurred while the spacecraft was located on the dayside (SZA $<$ 63$^\circ$) and close to its periapsis (altitude $h <$ 400~km).
Sounding pulses with frequencies $f$ greater than $\sim$1.1~MHz were found to be ducted, corresponding to $\sim$3 times the local plasma frequency $f_{pe}$ at the spacecraft in each case, and $\sim$250-1000 times the local cyclotron frequency $f_{ce}$.
\cite{zhang16a} argue that these ducts are artificially generated (as opposed to naturally occurring) field-aligned structures, formed in the Martian ionosphere by the relatively high power MARSIS sounder (the radiated energy being large compared to the typical thermal energy content of the local plasma).
In the case of similar artificial field-aligned irregularities responsible for ducting in the Earth's ionosphere, the required density cavity grows in the field-aligned direction in response to the sounding pulses when the ratio $f_{pe} / f_{ce}$ is close to a (low) integer value~\citep{benson97a}.

In this paper, we further examine one of the events first published by~\cite{zhang16a}, that of 2005 day 318 (MEX orbit 2359), being the most well-formed example of ionospheric ducting producing a so-called `epsilon' signature in several ionograms.
We investigate possible causal connections to another uncommon process in the Martian ionosphere -- the apparent extreme localized heating of the ionospheric electron plasma at altitudes close to the peak density.
This related process was first observed at Mars by~\cite{nielsen07b} in data from the same orbit, among others.

\section{Observations}

\begin{figure*}
    \centering
	\includegraphics[width=\textwidth]{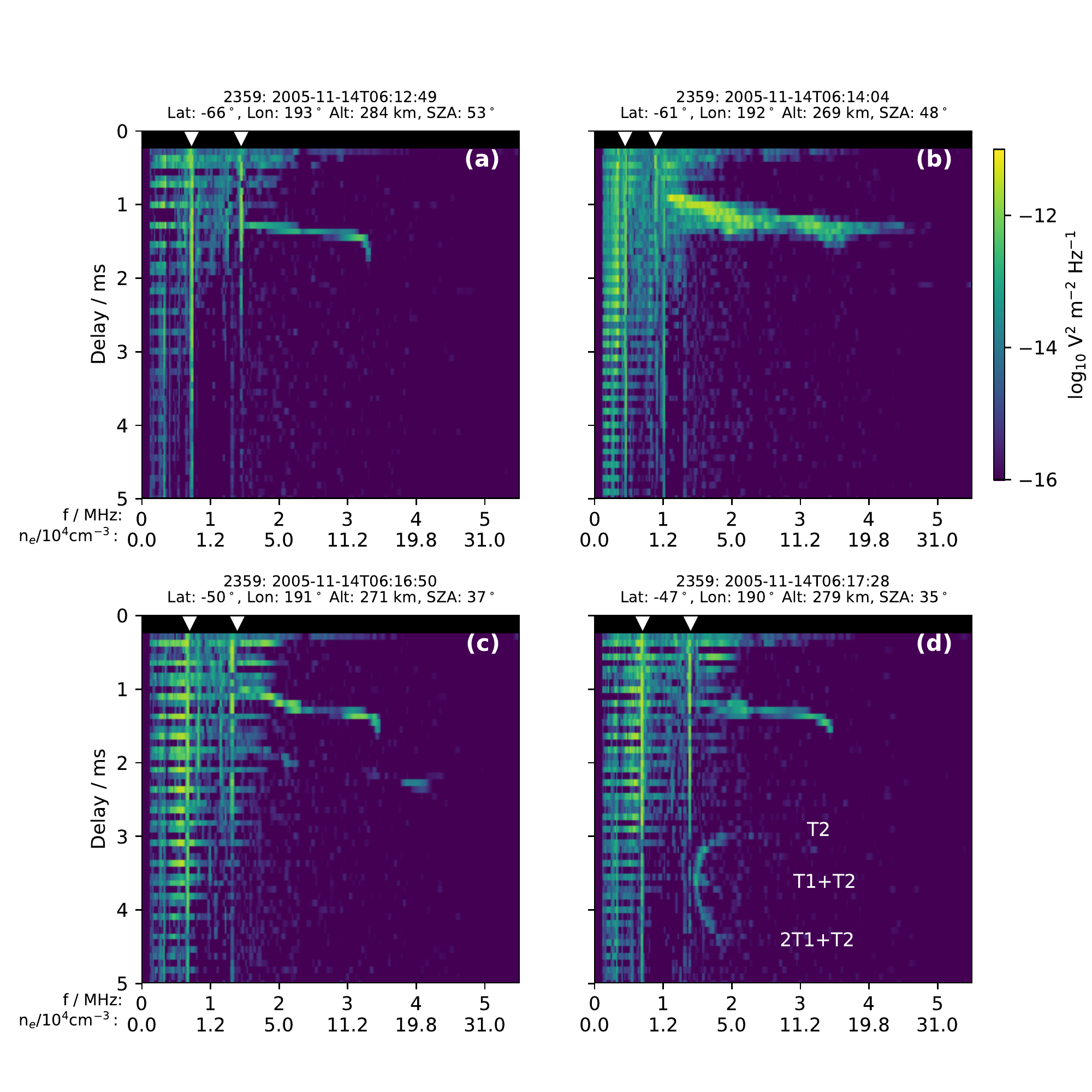}
	\caption{Four MARSIS ionograms obtained on orbit 2359. The received signal on the antenna is color-coded versus both delay time $\tau$ and sounding frequency $f$. Equivalent plasma densities are marked on the lower axis, and spacecraft position (latitude, longitude, altitude and solar zenith angle) in planetographic coordinates above the upper edge.
    White triangles along the upper edge indicate $f=f_{pe}$ and $f=2 f_{pe}$.
    }
	\label{fig:ionograms}
\end{figure*}

In Figure~\ref{fig:ionograms} we show four ionograms obtained on orbit 2359, over a period of $\sim$5 minutes close to periapsis at similar altitudes and longitudes, though with significant variation in both latitude and solar zenith angle.
The format of each is identical, with reflected intensity shown color-coded versus the sounding frequency $f$ and delay time.
Positional information for MEX is shown on the upper edge of each ionogram.
All four were taken on the dayside, and each shows a clear ionospheric reflection at a delay of $\sim$1-1.5~ms.
Each individual ionogram is acquired over an interval of $\sim$1.5~s.

Figure~\ref{fig:ionograms}a shows a fairly typical ionospheric reflection trace visible from $\sim$1.4 to $\sim$3.2~MHz, with the characteristic curve towards larger delays near the highest frequencies, indicating propagation of sounding pulses down to the ionospheric peak density (the so-called ``critical frequency'').
Numerical inversion of this trace using the technique described by~\cite{morgan13a} yields a peak altitude of approximately 130~km, in line with expectations based on statistical descriptions of the Martian ionosphere at this location~\citep[e.g.][]{morgan08a}.
The local plasma resonance at $\sim$0.7~MHz is clearly visible as a intense vertical line in the ionogram, and a second harmonic is also present.
From this we infer that the spacecraft is embedded in a plasma of density $n_e \approx 6500~\mathrm{cm}^{-3}$.
Similarly, several evenly spaced horizontal lines are present at low frequencies at multiples of the local electron gyroperiod, indicating a local magnetic field of $\sim$100~nT.
No surface reflection is visible in this or any of the following ionograms, likely due to the presence of a highly collisional low-altitude plasma layer somewhere below the main peak, in which even high-frequency waves are collisionally damped to the point that no reflected rays are able to be received.

The ionogram shown in Figure~\ref{fig:ionograms}b was  obtained at slightly lower altitudes and marginally closer to the sub-solar point than that shown in Figure~\ref{fig:ionograms}a.
The local magnetic field has increased significantly to $>$170~nT, as the spacecraft moved into a region of more intense crustal magnetic fields, while the local plasma resonance frequency dropped to $\sim$0.5~MHz, ($n_e \approx 3100~\mathrm{cm}^{-3}$).
Measurements of magnetic field strength by MARSIS are limited by the characteristics of the instrument, such that fields weaker than $\sim$5~nT or stronger than $\sim$190~nT cannot be resolved due to aliasing.
The ionospheric reflection is markedly different than that observed in Figure~\ref{fig:ionograms}a, just northward of the location of the spacecraft in Figure~\ref{fig:ionograms}b.
Firstly, the maximum frequency of the ionospheric trace has significantly increased to $\sim$4.5~MHz ($n_e \approx2.4\times10^5$~cm$^{-3}$).
This indicates a peak plasma density in the ionosphere well above the $\sim$1.3$\times10^{5}\mathrm{cm}^{-3}$ expected at this solar zenith angle according to models of the `nominal' ionosphere~\citep[e.g.][]{morgan08a}.
Furthermore, the lack of any significant curvature in the trace at the highest frequencies prevents us from making safe conclusions about the maximum plasma density in this location, and the true value could well be higher even than given.
Here, the ionospheric trace is everywhere more spread out in delay than the sharp nadir reflection shown in Figure~\ref{fig:ionograms}a, indicating a very non-planar ionosphere.
This spreading takes place over $\sim$0.5~ms in delay, corresponding to $\sim$75~km in apparent range, although the typical scales of the associated irregularities could well be much smaller owing to dispersive effects.
Some commonality is apparent with the `spread-F' signatures seen often in Earth's ionosphere, as large-scale (compared to the sounding wavelength) density structures scatter the reflected rays along multiple paths.

\cite{nielsen07b}, in an analysis of the iongram shown in Figure~\ref{fig:ionograms}b and other similar features observed on the same orbit, suggested that these greatly enhanced peak densities could be caused by a two-stream (Farley-Buneman) instability acting in the ionosphere.
Ionospheric plasma flows, driven ultimately by the action of the solar wind, would set up such two-stream interactions when the magnetized electrons are deflected upon flowing into a region of intense, irregular crustal fields, while the heavier ions are less perturbed.
Ultimately, this process is suggested to locally heat the electron plasma and therefore reduce the recombination rate, establishing a higher overall plasma density, and could furthermore provide a source for the density fluctuations evidenced by the delay-spread reflections.

Figure~\ref{fig:ionograms}c was obtained shortly after a second such peak density enhancement and delay spreading event.
The nadir trace from the ionosphere immediately below the spacecraft has now returned to a much more sharp reflection, without any appreciable degree of delay-spreading.
Meanwhile, the local plasma density has also risen, and is now broadly consistent with that in Figure~\ref{fig:ionograms}a.
However, a further trace is also discernible below the principal nadir reflection, extending still to high frequencies.
We suggest that this additional trace is associated with the peak density enhancement, having effectively ``detached'' from the original feature and moved to higher delays in the intervening soundings, appearing now in this ionogram at delays of $\sim$1.9-2.4~ms.
This is consistent with the source of the reflection being stationary in the ionosphere as the spacecraft recedes from it.
The enhanced density trace is in this case patchy, still does not display a clear curvature at its highest frequencies, and the maximum frequency of this trace continues to extend well above the peak frequency of the nadir trace.

This configuration of two distinct traces in a single ionogram bears some similarity to the so-called `oblique echoes' associated with ionospheric upwellings, as studied e.g.\ by~\cite{duru06a}.
This enhanced density trace does indeed arrive at the spacecraft at oblique incidence, and its variation along the orbit is also necessarily similar to the hyperbolic traces, as is the case for any reflection from a target fixed in the ionosphere.
However, its significantly elevated frequency seems to reliably rule it out as simply due to an otherwise unremarkable ionospheric upwelling of the type extensively studied previously.
Furthermore,~\cite{andrews14a} have demonstrated that the echoes associated with ionospheric upwellings are highly repeatable, being detectable in all orbits that cross over the same region.
Their relative stability and repeatability is in contrast to these peak density enhancements, which are only very rarely observed~\citep{zhang16a}.

The final ionogram shown in Figure~\ref{fig:ionograms}d depicts one of the so-called `epsilon' signatures, reported first in the Martian ionosphere by~\cite{zhang16a}.
In addition to a more-typical ionospheric reflection similar to that seen in Figure~\ref{fig:ionograms}a, three connected traces are are also visible over much larger delays of $\sim$3-4.4~ms.
These branches of the epsilon trace are labelled `T2', `T1+T2' and `2T1+T2' in common with the terminology used by~\cite{zhang16a}.
\cite{zhang16a} suggest that this particular signature is the result of ducted propagation of the MARSIS sounding pulse in a plasma cavity, by analogy to similar signatures noted in Earth-orbiting topside ionospheric sounders~\citep[e.g.,][]{dyson78a}.
By their estimation, this signature is persistent for four consecutive ionograms ($\sim$30~s), as the spacecraft moves through the density structure responsible for the ducting.
In this instance, the epsilon signature is visible at frequencies from $\sim$1.4 to $\sim$2.1~MHz, i.e. from about twice the local plasma frequency to about half of the maximum plasma frequency observed at the ionospheric peak.
At the spacecraft, the magnetic field is close to horizontal, still intense and dominated by crustal sources, and any field-aligned density cavity would therefore be expected to extend almost horizontally away from the spacecraft.
Ducting of the wave along these closed field lines provides an opportunity for  reflections to be detected from both single and multiple `hops' through this cavity (i.e., pulses that have been reflected from the ends of the cavity two or more times, returning to the spacecraft with sufficient intensity within the receiving interval of the sounder, as discussed e.g. by~\cite{zhang16a}).
In this example, the trace labelled `2T1+T2' is associated with a pulse that has traversed the paths T1, T2 and T1 again, before being received on the antenna.

Instances of the unusual ionospheric signatures shown in Figure~\ref{fig:ionograms}b and d are apparently rare in the MARSIS data set.
\cite{zhang16a} report a total of 7 observations of ducted wave propagation in an analysis of $\sim$8 years of data, with the example shown here in Figure~\ref{fig:ionograms}d being one of the clearest such detections.
Meanwhile, the significantly enhanced densities and irregularities associated with the Farley-Buneman instability were observed by~\cite{nielsen07b} on a total of three orbits from the first year of MARSIS operations.
That they both occur, in their respectively clearest and most extreme examples, on the same orbit and within a matter of minutes of each other clearly warrants further investigation.

\begin{figure*}
    \centering
	\includegraphics[width=\textwidth]{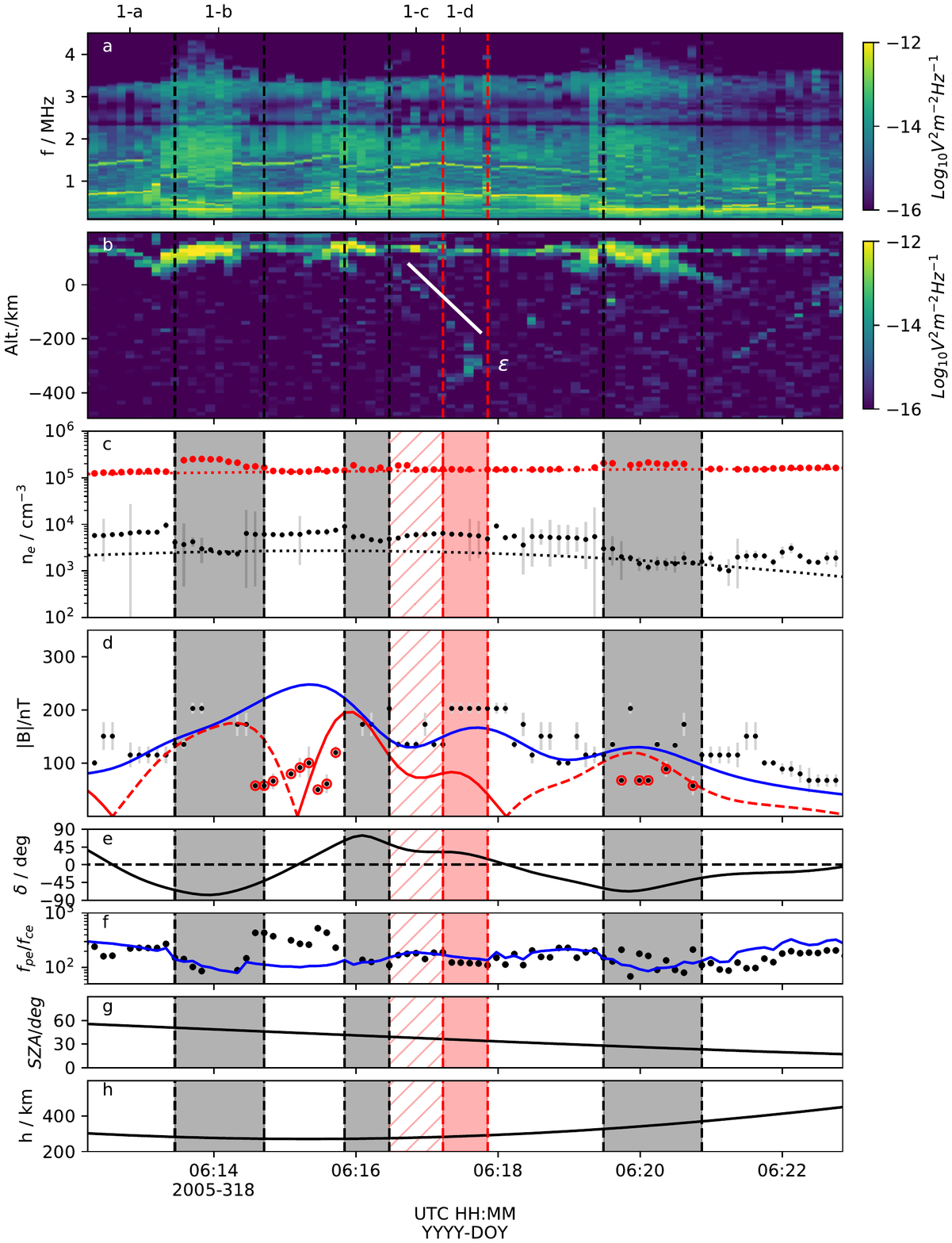}
    \caption{Caption next page.}
\end{figure*}
\addtocounter{figure}{-1}
\begin{figure*}
	\caption{Timeseries plots of MARSIS data obtained on 2005-318 (14 November, orbit 2359).
    a) Integrated reflected signal received color coded as a function of sounding frequency $f$ and time (a so-called `spectrogram').
    Tick marks on the upper edge indicate the timing of the ionograms displayed in Figure~\ref{fig:ionograms}.
    b) Reflected signal at a constant frequency ${f=1.9}$~MHz versus apparent altitude $h$ (a so-called `radargram').
    c) Local plasma density measured at the spacecraft altitude (black points, with vertical bars indicating uncertainty), and peak ionospheric plasma density (red points) determined from the maximum frequency of the ionospheric reflection.  Correspondingly colored dotted lines give expected nominal values at this location, to guide the eye in accounting for variations due the spacecraft motion in altitude and SZA.
    d) Magnetic field magnitude measured at the spacecraft (black points) by MARSIS. Points with a surrounding red circle are likely erroneous, aliased values. The blue line shows the expected crustal magnetic field magnitude at the location of the spacecraft according to the model of~\cite{cain03a}, while the red line gives the magnitude of the radial component of the field (dashed negative) from the same model.
    e) Magnetic zenith angle $\delta$, with ${\delta=0^\circ}$ indicating horizontal fields, and $\delta=\pm$90$^\circ$ vertical fields (positive upward).
    f) Black points show the ratio of the plasma and cyclotron frequencies $f_{pe}/f_{ce}$ at the spacecraft, computed from the black points in panels c and d. The blue trace instead uses the modelled magnetic field magnitude in the calculation.
    g, h) spacecraft solar zenith angle and planetographic altitude $h$.
    Intervals shaded grey in the lower panels, bounded by black dashed lines indicate regions of peak ionospheric plasma density increases, as studied first by~\cite{nielsen07b}. The red shaded and hatched region indicates the interval for which the epsilon feature was observed.
    }
	\label{fig:ts}
\end{figure*}

In Figure~\ref{fig:ts} we show all data obtained by MARSIS during the relevant segment of orbit 2359.
Timings of the four individual ionograms shown in Figure~\ref{fig:ionograms} are indicated by the labels on the upper edge of the figure.
MARSIS data is shown the format of a `spectrogram' in Figure~\ref{fig:ts}a, in  which the individual ionograms are processed by integrating the received signal at a given frequency over all delays.
The resulting plot of signal intensity as a function of sounding frequency is color-coded at plotted versus time along the orbit.
So processed, the local plasma resonances (vertical lines in Figure~\ref{fig:ionograms}) are visible here as horizontal lines, with spacing changing with time as the local plasma density changes along the orbit.
Figure~\ref{fig:ts}b displays a MARSIS `radargram', i.e.\ a cut through the ionograms at fixed frequency $f=1.9$~MHz, plotted versus apparent altitude instead of delay.
Here, the intense horizontal line at $\sim$100~km apparent altitude is the nadir reflection.
The lower panels c - h then show measured local and peak electron densities, magnetic field intensity, magnetic inclination angle, the ratio $f_{pe}/f_{ce}$, and the SZA and altitude of MEX.
Throughout the interval shown, the spacecraft moved to progressively lower solar zenith angles, and periapsis was reached at 06:15.
Vertical red dashed lines and shading through each panel span the period in which the epsilon signature was reported by \cite{zhang16a}, while vertical black dashed lines and grey shading show the three peak density increases noted by~\cite{nielsen07b} on this orbit.

The time variation of the peak frequency of the ionospheric reflection can be discerned by examining the highest frequency signals in Figure~\ref{fig:ts}a which rise above the dark blue background intensity.
Within the three grey shaded intervals, the peak frequency significantly increases, indicating increases in peak plasma density (panel c, red points), along with the delay spreading leading to a broader signature at the same time in panel b.
Each of these peak density increases is observed in regions of near-radial crustal fields, evidenced by the dominant contribution of the radial field to the total modelled value (panels d and e).
The peak density increase is significant compared to the measurement accuracy, as seen from the generally close agreement between the measured and expected peak densities (panel c, red points).
The first and last peak density increases marked on the figure are the most significant in both amplitude and duration, while the central event starting just prior to 06:16 is both shorter in duration, and with a less pronounced density increase.
It can be clearly seen that the local plasma density in each case decreases during these intervals, with the black points in panel c dropping by up to a factor of $\sim$5 compared to values at the approximate boundaries of these intervals.
This fact was not reported by~\cite{nielsen07b}, and indicates a major reconfiguration not just of the deep ionosphere close to the peak, but also a significant reduction in the effective scale height of the topside ionosphere above this.

Within the presentation format of Figure~\ref{fig:ts}a, the epsilon signature seen in Figure~\ref{fig:ionograms}d is not visible as the intensity of its three branches is comparable to or weaker than the nadir ionospheric reflection, and does not therefore contribute significantly to the log-scaled intensities shown.
However, in panel b, the cut through the received signal at $f=1.9$~MHz shows the epsilon traces as a cluster of bright `pixels' at the point marked by the white $\epsilon$ character.
As was reported by~\cite{zhang16a}, the branches of the epsilon signature  converge with time, and the whole structure apparently recedes from the spacecraft and moves to higher frequencies.
Only the upper `T2' branch of the epsilon is clearly persistent in Figure~\ref{fig:ts}b, and we have placed a white diagonal bar directly above it to guide the eye.
While its intensity remains weak compared to the nadir reflection throughout the period, we nevertheless suggest that the reflections discernible below the white bar are attributable to the same single structure, extending well beyond the interval suggested by~\cite{zhang16a}, clearly emerging from the nadir reflection at $\sim$06:16:15.
We indicate the extension of the interval for which a trace associated with the epsilon signature is visible by the red hatched region in~\ref{fig:ts}.
Significantly, this leads to a clear connection between the epsilon signature reported by~\cite{zhang16a} and the peak density enhancements reported by~\cite{nielsen07b}, with the final reflection site of the T2 branch of the epsilon trace (the ionospheric footprint of the ducting field line) being co-located with the peak density enhancement.

In order to investigate the configuration of the crustal magnetic field throughout this periapsis pass, we display the magnetic zenith angle $\delta$ in Figure~\ref{fig:ts}e.
Purely vertical (cusp-like) fields have $\delta=\pm90^\circ$, while horizontal fields have $\delta=0^\circ$.
It is readily apparent that the interval for which the epsilon signature is  detected occurs in fields that are closer to horizontal than vertical, the average value of $\delta$ during the red shaded interval being $\sim$25$^\circ$.
Furthermore, we note a monotonic decrease in $\delta$ from near-vertical fields at the spacecraft at the time when the immediately preceding peak density enhancement was detected, to the near-horizontal fields when the ducted echoes observed.
This smooth variation of $\delta$ from values of $\sim$70$^\circ$ to $\sim$15$^\circ$ indicates that throughout this period, $\sim$06:15 to $\sim$06:18, MEX traversed a single `arcade' of magnetic fields, including a single `cusp', above which the peak density enhancement was detected.
Further towards the sub-solar point, but still on the same crustal field lines, the ducted sounding pulses were later observed.
While the crustal field also does vary somewhat in the cross-track direction, it does generally display a high degree of zonal symmetry in this region, such that the dominant variations in both direction and intensity are in the north-south direction, and more closely aligned with the spacecraft orbital trajectory.
Thus, while MARSIS does not resolve angle of incidence directly for received reflections, we are nevertheless confident that the most significant features present in the data on this orbit are likely to be reflections from locations close to the trajectory.

As the spacecraft moves through the ionospheric irregularities, from a region of near-vertical to near-horizontal crustal fields, the received signals transform seamlessly from the high-frequency, delay-spread echo seen in Figure~\ref{fig:ionograms}b, to the multiple ducted traces seen in Figure~\ref{fig:ionograms}c-d.
Meanwhile, a nominal nadir reflection is uniformly visible once the high-frequency echo is no longer observed.
That the individual branches of the epsilon trace seen in Figure~\ref{fig:ionograms}d do not also undergo delay-spreading in the same manner as the high frequency echo shown in Figure~\ref{fig:ionograms}b indicates that the ducted pulses are propagating parallel to the field within a narrow density cavity.
Scattering from the density irregularities, which produced the delay-spread echoes no longer occurs when the irregularities instead duct the sounding pulses along the field.

Throughout the interval shown in Figure~\ref{fig:ts}, local plasma oscillations are also produced and detected by the sounder, visible as the most intense quasi-horizontal lines in panel a, at frequencies of $\sim$0.2-0.9~MHz.
From these, the local plasma density can be determined as shown by the black circles in Figure~\ref{fig:ts}c.
Generally these are elevated with respect to their long-term averages at the same location, according to the model of~\cite{andrews15a}, shown by the dotted black line.
Throughout this orbit, the modelled values lie significantly below the data, although the along-orbit variation is well represented.
The underestimated values are likely due to seasonal effects not present in the simple model.
No clear drop in local electron density is present during the interval highlighted by~\cite{zhang16a}, although it does end with an abrupt increase in density for a single ionogram.
Instead, in our estimation the signatures of the epsilon structure extend to marginally earlier times, beyond those reported by~\cite{zhang16a}, to the start of a (brief) peak density enhancement reported by~\cite{nielsen07b} at $\sim$06:15:50.
This does appear to coincide with the spacecraft crossing a sharp local density gradient, and moving through a local density cavity from $\sim$06:15:50 to $\sim$06:17:50, with a relative depletion of $\sim$10-30\%.
However, only a few \% depletion is required to support ducted propagation and the production of the epsilon signature, on the basis of related observations from Earth-orbiting topside sounders~\citep[e.g.][]{muldrew63a}.

\begin{figure*}
  \centering
  \includegraphics[width=\textwidth]{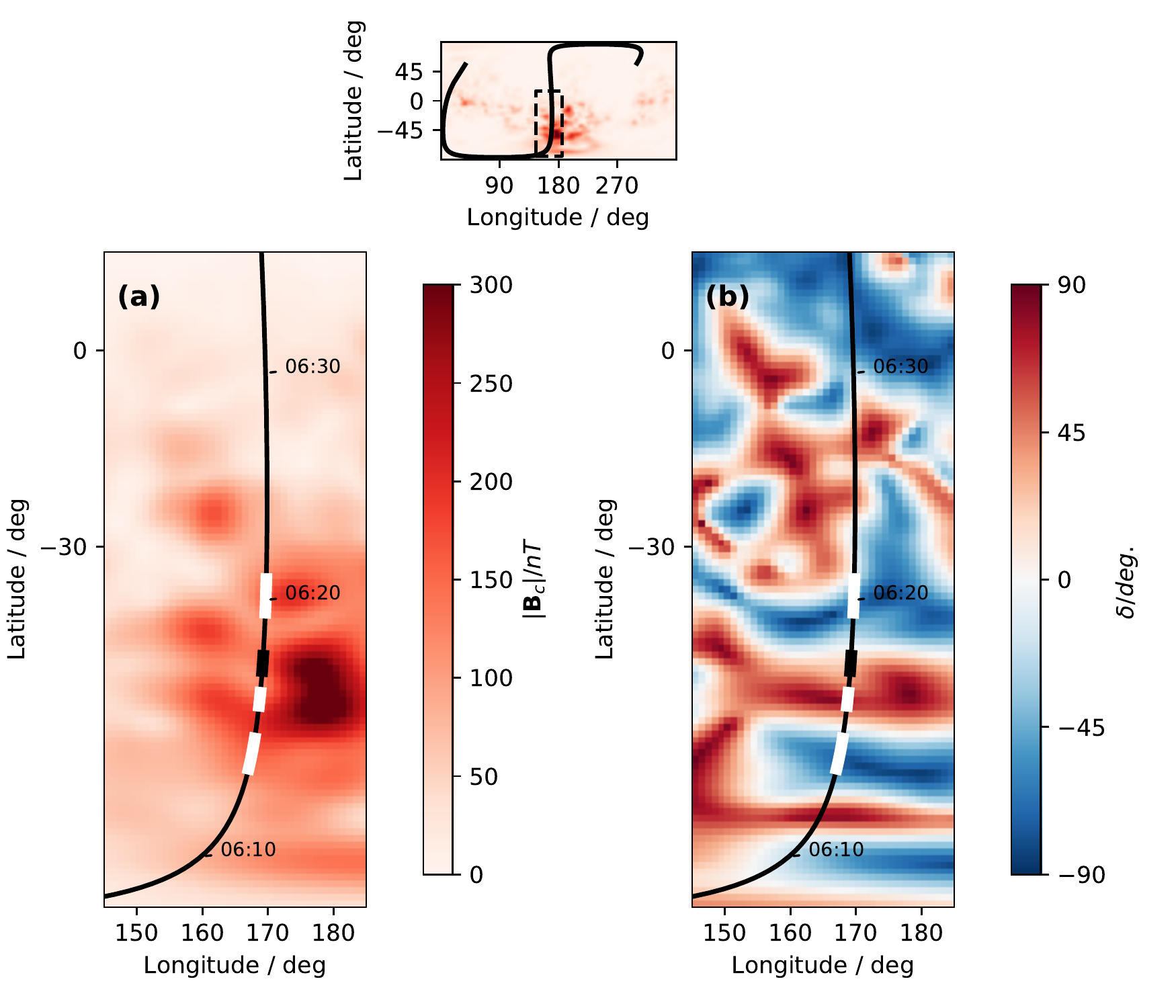}
  \caption{Projections of the orbit of MEX onto the surface of Mars.
  The two large panels show the MEX orbit trajectory plotted versus planetographic latitude and longitude (black line), with the intervals of peak density enhancements highlighted by the white blocks, and the interval for which the epsilon signature is clearly visible by the black block. The magnitude $|B|$ of the Martian crustal field is shown underneath the orbit in panel (a), while the inclination angle $\delta$ is shown in panel (b). Both parameters are determined from the~\cite{cain03a} model, evaluated at a fixed 300~km altitude.  The upper inset shows the larger context of the two main panels.}
  \label{fig:map}
\end{figure*}

The correspondence between the observed ionospheric signatures and the crustal field structure is further depicted in Figure~\ref{fig:map}, in which the orbit trajectory of MEX is projected onto the surface of Mars.
Figure~\ref{fig:map}a and b also show the crustal field magnitude $|B|$ and inclination angle $\delta$, respectively, calculated again from the~\cite{cain03a} model at 300~km altitude (corresponding approximately to the altitude of MEX during the interval).
The peak density enhancements, highlighted by the three white segments are seen to occur in regions with near-vertical fields (large positive or negative $\delta$).
The epsilon signature meanwhile is observed on the northern edge of one such longitudinally extended vertical field region, in the transition to near-horizontal fields.

No other complete epsilon signatures are observed on this orbit, although several intermittent distant reflections are observed following the other peak density enhancements originally reported by~\cite{nielsen07b}.
Additionally, occasional oblique reflections at frequencies between $f_{pe}$ and $f_{max}$ are also present, for example the hyperbolic traces visible in Figure~\ref{fig:ts}b from $\sim$06:19:00 to $\sim$06:21:30, and separately from $\sim$06:21:30 to the end of the displayed interval.
These are commensurate with the type of oblique reflections previously reported e.g.\ by~\cite{duru06a}, attributed to stable ionospheric upwellings.
We conclude therefore that the processes that give rise to both the peak density enhancements and the ducted echoes are present not just at a single location in Mars's ionosphere during this period, but at several locations with favourable magnetic field geometry. Furthermore, they occur in addition to other more typical and widely reported modifications of the ionosphere by the presence of intense crustal fields.

\cite{zhang16a} suggest that the density depletion that gives rise to the ducted propagation and resulting epsilon signature may be artificial in origin, with the high power MARSIS sounding pulse rapidly establishing a field-aligned plasma density depletion through the ponderomotive force.
This mechanism has been suggested to be effective in certain situations when sounding the Earth's ionosphere with similar instruments~\citep{benson97a}, in the case where the local plasma frequency $f_{pe}$ is a low integer multiple of the cyclotron frequency $f_{ce}$, $f_{pe}/f_{ce} \approx n$ for integer n.
\cite{benson97a} discuss various examples for which $n=3$ and $n=4$, but note that a full theoretical description of the coupling is lacking.
However, as can be seen in Figure~\ref{fig:ts}e, for this event at least the period in which the epsilon is most clearly observed has $f_{pe} / f_{ce}$ $\sim$100-1000, as first reported by~\cite{zhang16a}, somewhat stretching the postulated requirement $f_{pe}/f_{ce} \approx n$.
While the measurement of the local cyclotron frequency by MARSIS is increasingly unreliable at these $>$100~nT values (and clearly aliased measurements are present, as indicated by the red circles in Figure~\ref{fig:ts}d), we nevertheless obtain very similar values for $f_{ce}$ when instead using a model of the crustal magnetic field strength at the location of the spacecraft, as seen by the blue trace in Figure~\ref{fig:ts}d.

\section{Summary and Discussion}

Two unusual signatures have been noted in the MARSIS data obtained on MEX orbit 2359, during day 318 of 2015.
Firstly, significant apparent enhanced peak plasma densities at $\sim$130-150~km altitudes and associated evidence for large-scale density irregularities are noted at three distinct locations on the orbit as the spacecraft passes through regions of near vertical crustal magnetic fields (e.g., Figure~\ref{fig:ionograms}b).
\cite{nielsen07b} suggest these are due to the localized heating of the ionosphere near the peak, possibly as the result of unstable flows resulting ultimately from solar wind driving.
Secondly, shortly after one such peak density enhancement, ducted propagation of the MARSIS sounding pulses is observed, indicating the presence of field-aligned density cavities at spacecraft altitudes, in near-horizontal fields (e.g., Figure~\ref{fig:ionograms}d).
Here, we have shown that these two phenomena appear linked by continuous signatures in the data, and take place on the same or similar field lines (noting that the zonal crustal field in this region is relatively weak, but can still lead to an East-West displacement in the footprints of these field lines in the ionosphere).

\begin{figure*}
    \includegraphics[width=\textwidth]{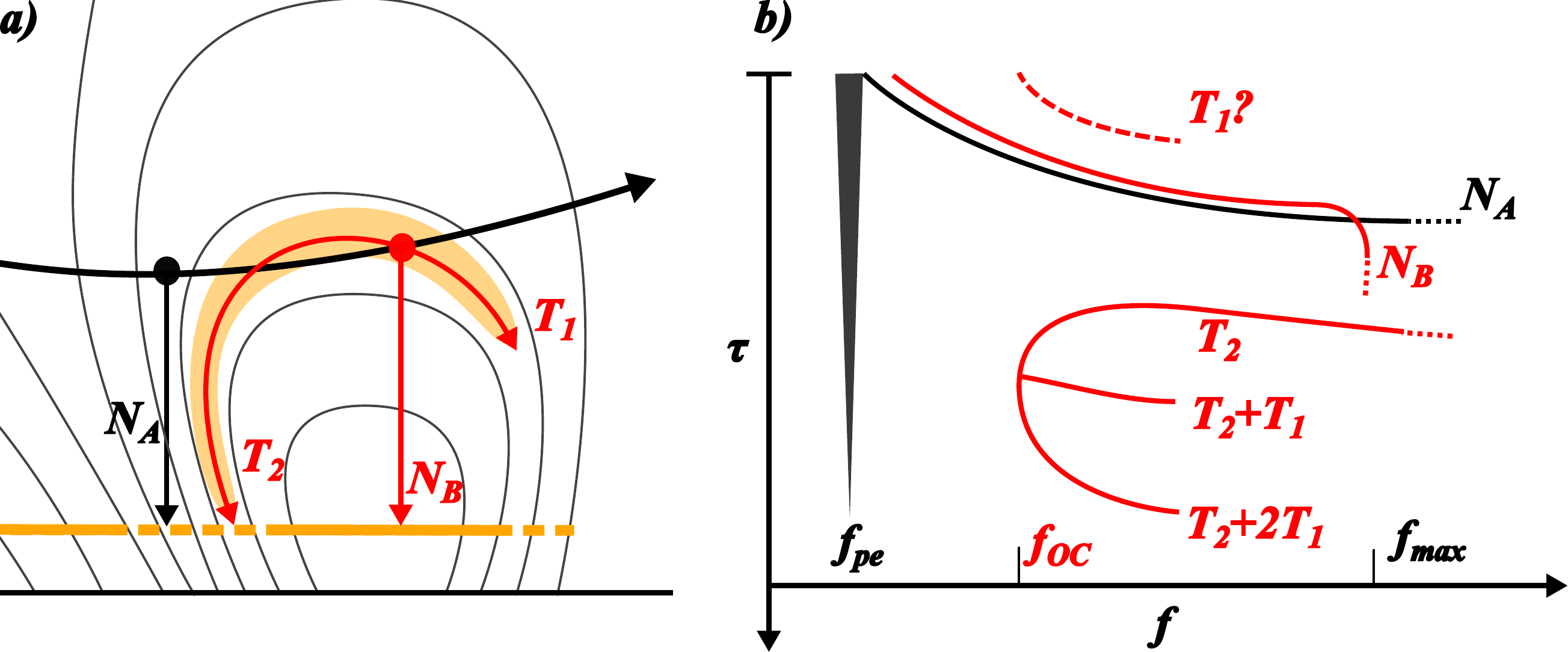}
    \caption{a) MEX's trajectory and MARSIS radar echoes obtained at two locations during a passage through a crustal field arcade.
    Thin black lines depict the crustal field, the thick black line MEX's trajectory. Black and red circles indicate the location of MEX during two MARSIS soundings, A and B.  Arrowed lines of the same color indicate the path of radar reflections, labeled $N_A$ and $N_B$ for the two nadir reflections and $T_1$ and $T_2$ for the two ducted traces seen in sounding B.  The horizontal solid orange line indicates the altitude of the ionospheric peak, drawn dashed in the region where enhanced peak densities are observed.  Meanwhile, the orange shaded region schematically indicates the structure of the ducting region along the field lines.
    b) Illustrative features of the two ionograms corresponding to the locations in a), versus time delay $\tau$ and sounding frequency $f$. The same color code applies for the two ionograms.  The resonance `spike' at the local plasma frequency $f_{pe}$ is present in both cases (harmonics $n f_{pe}$ omitted for clarity). $f_{OC}$ indicates the low frequency cutoff of the ducted traces.}
    \label{fig:eg}
\end{figure*}

In Figure~\ref{fig:eg} we sketch our interpretation of this event, showing qualitatively the physical ionospheric and sounding geometry (a) and the resulting reflections received by MARSIS (b).
Prior to the arrival of the spacecraft an initial density perturbation was formed in the deep ionosphere, perhaps as the result of the ionospheric flows suggested at low altitudes by~\cite{nielsen07b}, indicated by the dashed portion of the orange horizontal line depicting the altitude of the ionospheric peak.

As soundings are performed in near-vertical field regions (at the black circle), the nadir reflected trace is spread over a broad range of delay, as well as showing the highly enhanced peak densities described by~\cite{nielsen07b}.
Simultaneously, density perturbations are produced along crustal field lines to higher altitudes.
Whether these consist of depletions or enhancements, the resulting structure in the refractive index results in a natural radio duct indicated by the orange shaded region.
Then, as the spacecraft moves into more horizontal fields (at the red circle), a relatively `clean' nadir reflection is instead observed from the unperturbed ionosphere below the spacecraft.
This is in addition to more complex signatures simultaneously detected resulting from ducted propagation of the sounder pulses along the field-aligned irregularities, including a long-distance duct that leads to reflections from the ionospheric peak, close to the region sampled by the nadir case from the first sounding.
The most obvious signature of these ducted traces is the `epsilon' noted by~\cite{zhang16a}.
We note the apparent lack of the isolated corresponding `T1' trace in Figure~\ref{fig:ionograms}d, which is expected at delays smaller than that of the nadir reflection on the basis of the properties of the `T2' and `T1+T2' traces (from which the expected `T1' trace can be uniquely determined).
This missing `T1' trace may simply be the result of the intense interference at low delays and low frequencies associated with the localised antenna-plasma interaction.
We also note that both the nadir and ducted reflections that take place in the regions of enhanced peak plasma density appear to be gradually attenuated toward higher frequencies (Figure~\ref{fig:ionograms}b), rather than having a sharp maximum frequency above which they are no longer detected.
This could indicate that the true density could well be higher than those estimated here.

\cite{zhang16a} suggest that the low-frequency cutoff of the epsilon signature, i.e.\ the point at which the three branches merge to a single point, is related to the artificial formation mechanism, although the theory first put forward by~\cite{benson97a} requires a highly magnetized plasma, unlike that present at Mars.
We suggest that it instead is determined purely by the field-perpendicular gradients in the density cavity (or, smaller sub-cavities).
The efficiency with which the duct is able to guide waves will be strongly dependent on frequency, and sensitive to the magnitude and gradient in the electron density.
Small variations in these parameters will change the effective `opening-angle' of the duct, and lead to changes in the ducted frequencies~\citep[see e.g.][]{calvert95a}.
In this case, the low frequency cutoff of the duct is then purely a result of the spatial structure of the plasma density within duct, and the varying location of the spacecraft within it.

If the density irregularities responsible for the ducted propagation are found at all altitudes down to the ionospheric peak on the same field lines, then it is reasonable to suggest that the delay-spread nadir reflections observed earlier on the orbit could be the result of the same irregularities.
We can estimate the perpendicular length scale of the irregularities responsible, as the spacecraft is apparently contained within the same ducting cavity for at least the time taken for a single MARSIS ionogram to be obtained ($\sim$1.5~s of actual sounder operation).
Assuming a horizontal field for the duration of the observed epsilon signature, and perpendicular (vertical) velocity of MEX 0.5~$\mathrm{km\;s^{-1}}$ yields a minimum perpendicular length scale of $\sim$750~m.
Density cavities with smaller length scales would give rise to intermittent ducted echoes as the spacecraft enters and exits individual density structures during the course of a single sounding, not consistent with observations.


Meanwhile, estimates of the spatial extent of the cavity along the field line can be obtained by considering the observed time delay of the different branches.
For the `T2' and `T1+T2' branches depicted in Figure~\ref{fig:ionograms}c, these maximum delays are $\sim$3.1 and $\sim$3.7~ms, respectively.
Utilising the same inversion scheme employed for the nadir reflection described by~\cite{morgan13a}, the longer (T2) duct branch can be tentatively estimated to be $\sim$300-400~km in length.
While this method may not be strictly appropriate in this situation, we nevertheless are satisfied that it yields at least a reasonable order of magnitude estimate for the extent of the duct.
These are then comparable to the estimated length of the field lines encountered by the spacecraft at this location, again suggesting that the irregularities responsible for the ducting can be present down to the peak of the ionosphere.

Recently, MAVEN observations have shown evidence of ionospheric irregularities produced by the Farley-Buneman instability acting in the lower Martian ionosphere~\citep{fowler17b}, analogous to the E-region of Earth's ionosphere.
In-situ measurements by instruments on MAVEN show fluctuations in both density and magnetic field in regions of near-horizontal fields with similar characteristic scales as those inferred for the ducting field-aligned irregularities studied here.
In this instance, there can be no question that these are a natural feature of the Martian ionosphere, as while the Langmuir Probe instrument on MAVEN does act as a sounder, transmitting a low power white noise signal through the booms on which the probes are supported, it does so with less than 0.1\% of the effective radiated power of MARSIS.
\cite{fowler17b} report density variations of up to $\sim$200\%, much greater than the few \% typically required to lead to ducted wave propagation in the Earth's ionosphere.

For ducted echoes observed by topside sounders at Earth, it is observed that the process is much more efficient for the X-mode rather than the O-mode, with the `opening-angle' of the duct being significantly smaller for the O- than the X-mode~\citep{muldrew69b}.
This may explain in part why ducted echoes are only rarely observed by MARSIS, as the much weaker typical magnetic field strength encountered at Mars, even in regions of relatively intense crustal fields are so low that O-~and X-mode effects are rarely, if ever, relevant.
We also note again that radar spread-F is also routinely associated with the presence of ducted echoes~\citep{muldrew69b}.

Following~\cite{nielsen07b}, we suggest that intense electric fields at lower altitudes near the ionospheric peak are the most likely cause of the observed peak density enhancements, density irregularities, and ducted propagation of the sounding pulses.
Such electric fields could conceivably be generated due to winds in the neutral atmosphere, external forcing by the solar wind, gradients in ionospheric conductivity, among other sources.
For sufficiently strong electric fields, the relative drift between partially demagnetized ions and magnetized electrons can exceed the local sound velocity, causing the onset of plasma turbulence and electron heating.
At these altitudes, molecular ions dominate, and are lost by dissociative recombination, the rate coefficient of which decreases with increasing electron temperature~\citep[see e.g.][]{schunk04a}.
Thus, the net effect is a rapid and localized increase in plasma density, as suggested by~\cite{nielsen07b}, with this situation persisting so long as the driving electric field is maintained.

Additionally we suggest that the same intense electric fields should lead instead to a decrease in plasma density at higher (spacecraft) altitudes, as seen in the data presented in this paper.
Here, friction between the (now magnetized) $\mathbf{E}\times\mathbf{B}$ drifting ions and the neutral atmosphere leads to ion heating, while having little effect on the electron temperature.
At higher altitudes near the spacecraft, atomic ions dominate.
The ion loss rate is controlled by the rate of charge exchange with neutral species to form molecular ions, which then quickly dissociatively recombine.
Increased ion drift speeds in these regions increase the cross-section for charge exchange and hence decrease the plasma density, as observed.
Qualitatively similar processes are widely studied in the Earth's F-region (upper) ionosphere~\citep[e.g.][]{killeen84a}.

The observed ducted propagation of the MARSIS sounding pulses suggests a spatially structured plasma in the direction perpendicular to $\mathbf{B}$, and this is expected to reflect the structure of the electric field also.
The associated enhanced plasma drift should therefore be mostly aligned along approximately constant magnetic flux, i.e. mostly in the East-West direction.
Following the discussion above, we suggest that these drift channels would have widths of $\sim$750~m, although the East-West extent would be much larger and not resolved by these (single spacecraft) observations.

In summary, recently analysed MARSIS data further suggests that Mars's highly structured crustal fields exert a significant influence on the structure of the ionosphere on the small scales required to affect the propagation of $\sim$MHz radio transmissions.
Crustal fields, able to modify ionospheric plasma flows driven by the solar wind interaction, can lead to large local plasma density enhancement, and the generation of field-aligned irregularities at all ionospheric altitudes.
Future observations with both MEX and MAVEN are expected to provide much more detailed understanding of these processes, along with more rigorous comparison with similar processes in the terrestrial ionosphere.

\section*{Acknowledgements}
Work at IRF was supported by grants from the Swedish National Space Agency (DNR 162/14) and the Swedish Research Council (DNR 621-2014-5526).
Work at Iowa was supported by NASA through contract 1224107 from the Jet Propulsion Laboratory.
Work at Boston University was supported by NASA award NNX15AM59G.
All data used in this paper are available in the ESA planetary science archive, \mbox{https://archives.esac.esa.int/psa}.

\end{document}